\documentclass[aps,prl,twocolumn,groupedaddress,10pt,showpacs]{revtex4-1}
\usepackage{graphicx}
\usepackage{amsmath,amssymb,bm}
\usepackage{url}
\def\figp#1#2#3#4{
\begin{figure}[!tp]
\begin{center}
\includegraphics[width=#3\textwidth,bb=#4]{./figures/#1.pdf}
\caption{#2}
\label{fig:#1}
\end{center}
\end{figure}}
\def\figpw#1#2#3#4{
\begin{figure*}[!tp]
\begin{center}
\includegraphics[width=#3\textwidth,bb=#4]{./figures/#1.pdf}
\caption{#2}
\label{fig:#1}
\end{center}
\end{figure*}}
\newcommand{\ply}{P_{L|Y}}
\newcommand{\pyl}{P_{Y|L}}
\newcommand{\PYL}{P_{Y\!L}}

\begin{document}

\title{Micro-Macro Relation of Production\\[5pt]
-- The Double Scaling Law for Statistical Physics of Economy --}

\author{Hideaki Aoyama}
\email[]{hideaki.aoyama@scphys.kyoto-u./ac.jp}
\affiliation{Department of Physics, Kyoto University, Kyoto 606-8501, Japan}
\author{Yoshi Fujiwara}
\affiliation{ATR Laboratories, Kyoto 619-0288, Japan}
\author{Mauro Gallegati}
\affiliation{ Dipartimento di Economia, Facolta di Economia ``Giorgio Fua'',
    Universit\`a Politecnica delle Marche, Piazzale Martelli 8, 60121 Ancona, Italy}

\date{\today}

\begin{abstract}
We show that an economic system populated by multiple agents generates an equilibrium distribution in the form of multiple scaling laws of conditional PDFs, which are sufficient for characterizing the probability distribution. The existence of the double scaling law is demonstrated empirically for the sales and the labor of one million Japanese firms. Theoretical study of the scaling laws suggests lognormal joint distributions of sales and labor and a scaling law for labor productivity, both of which are confirmed empirically. This framework offers characterization of the equilibrium distribution with a small number of scaling indices, which determine macroscopic quantities, thus setting the stage for an equivalence with statistical physics, bridging micro- and macro-economics.
\end{abstract}

\pacs{
02.50.-r, 
05.70.-a, 
89.65.-s, 
89.75.Da 
}

\maketitle

Economics is in crisis. Although there exists a mainstream approach
\cite{kydland1982time,gali2008monetary}, its internal coherence and ability in explaining empirical evidences are increasingly questioned. The causes of the present state of affairs go back to the mid of the XVIII century, when new figures of social scientist ({\it economists}) borrowed the method ({\it mathematics}) of the most successful hard science ({\it physics}) allowing for the mutation of {\it political economy} into {\it economics}. It was, and still is, the Newtonian mechanical physics of the XVII century, which rule economics. 

From then on, economics lived its own evolution based on the classical physics assumptions ({\it reductionism, determinism and mechanicism}). Quite remarkably, Keynes adopted the approach of statistical physics, which deeply affected physical science at the turn of the XIX century by emphasizing the difference between micro and macro, around the mid 1930s \cite{keynes1936general}. However, after decades of extraordinary success it was rejected by the neoclassical school around the mid 1970s, which framed the discipline into the old approach and ignored, by definition, any interdependence among economic agents (firms, banks, households) and difference between microscopic individual behavior and macroscopic aggregate behavior. 

The ideas of natural laws and equilibrium were transplanted into economics {\it sic et simpliciter}.  
As a consequence of the adoption of the classical mechanics paradigm, behavior of macro-economic systems are treated as a scaled-up version of one individual agent, who is called Representative Agent (RA), and complexity that emerges from aggregation was lost. 
Any learned physicist knows that this is entirely wrong for physical systems with many constituents: macroscopic behavior of gas is qualitatively different from that of a single molecule. Likewise, economy of a country is not explained by analysing a single RA as if he is on a deserted island all by himself, like Robinson Crusoe without even Friday, and multiplying the number of population to the results.

Another quite a dramatic example is the concept of equilibrium. In many economic models equilibrium is described as a state in which (individual and aggregate) demand equals supply. The notion of statistical equilibrium, in which the aggregate equilibrium is compatible with stochastic behavior of the constituents, is outside the box of tools of mainstream economists. Again, physics teaches us that the equilibrium of a system does not require that every single element be in equilibrium by itself, but rather that the statistical distributions describing macroscopic aggregate phenomena be stable.

What modern physics can do for economics is, then, to open a way to a proper treatise of macro economy as an aggregation of individual economic agents,
which is one main theme of econophysics on real economy \cite{gatti2006em,bible2010}.
Such a thought is not totally unfamiliar to open-minded economists,
whose keyword is {\it Heterogeneous Interacting Agents} (HIA) \cite{blume2006economy}, where `heterogeneity' implies that each has different characteristics; 
different financial profile like different energy and momentum,
and `interaction' is trade with exchange of money, goods, workers, information, etc.\ just as physical particles interact with each other.

In this letter we will show that a system populated by many HIA generates equilibrium distribution in the form of scaling laws. In particular, economic literature has shown the existence of large and persistent differences in labour productivity across industries and countries \cite{aoyama2008,ikeda2009iclpd,aoyama2008b}. Productivity is often measured in terms of the ratio between firms' revenues and the number of employees. It can be expected to be a unique value only within a very straightjacket hypothesis, such as the Representative Agent. If agents are heterogeneous and interact, then scaling laws emerge, and dispersion is nothing but a consequence of it. Using Japanese data we empirically demonstrate it. The conclusive remarks points out that a thermodynamical approach (see also \cite{foley1994statistical,aoki2006rmp}) may be what economics needs if HIA are the actors of the drama.

For the purpose to study properties in probability distributions, it
is essential to observe a large portion of the entire population of
firms and workers. A database of only listed firms, for example, is
insufficient to analyze properties of distributions. We employ the
largest database of Credit Risk Database (CRD) in Japan (years 1995 to
2009), which includes a million firms and fifteen million workers in
the year 2006, covering the large portion of the whole domestic
population. Below we give our analysis for the year 2006, but note
that the qualitative results are valid for other years as well.

We measure the value added $Y$ and the labor $L$ of each firm to have
the information of output and input in the production at the
individual level. We use simply the business sales/profits as a
proximity to the value added, and the end-of-year number of workers
(excluding managers) as the labor in order to calculate distributions
for $Y$ and $L$ and to uncover their properties.

To understand how workers are distributed among different levels of
output and productivity, we shall study the distributions of $Y$ and
$L$ using the following probability density functions (PDFs). The
joint PDF, $\PYL$, the conditional PDFs, $\pyl$ and $\ply$, and the
marginal PDFs, $P_L$ and $P_Y$, are defined by
\begin{equation}
\PYL(Y,L)=\pyl(Y|L)P_L(L)=\ply(L|Y)P_Y(Y).
\label{pjoint1}
\end{equation}
The conditional average of $f(Y)$ is defined by
\begin{equation}
E(f(Y)|L):=\int_0^\infty f(Y)\pyl(Y|L)dY,
\label{eyldef}
\end{equation}
for an arbitrary function $f(\cdot)$, and similarly for $E(f(L)|Y)$.

\figp{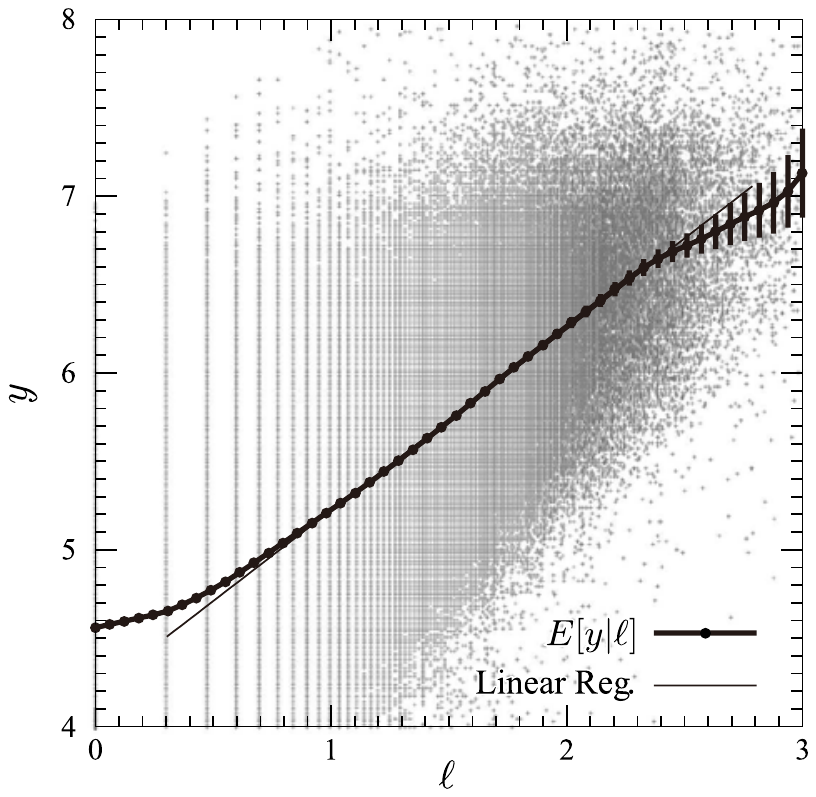}{%
  Scatter plot for $(L,Y)$ (gray dots). The curve with error bars is
  the nonparametric estimation for $E(y|\ell):=E(\ln Y|L)$, while the
  thin straight line is the linear regression (both axes are in units of
  $\ln 10$.)}{0.4}{188 315 418 540}

As we shall see, since the PDFs are heavy-tailed for $Y$ and $L$, it
is convenient for the purpose of statistical analysis to take the
logarithms of variables:
\begin{equation}
y:=\ln \frac{Y}{Y_0}, \quad \ell:=\ln \frac{L}{L_0},
\end{equation}
where $Y_0$ and $L_0$ are arbitrary scales.

Fig.~\ref{fig:fig1} shows the scatter plot for $(\ell,y)$. To
reveal statistical structure in data, which can be easily missed by
parametric methods, we employ a kernel-based nonparametric methods
\cite{li2007ne}. Fig.~\ref{fig:fig1} depicts a nonparametric
regression curve with error bars (95\% significance level) for
$E(y|\ell)=E(\ln Y|L)$. We can observe that there exists a range
$10^{0.7}<L<10^{2.5}$ for which the relation:
\begin{equation}
E(y|\ell)=\alpha\ell+\text{const.},
\label{eylalpha}
\end{equation}
holds where $\alpha$ is a constant. In fact, the goodness of fit for
nonparametric regression ($R^2=44.04\%$; see \cite{hayfield2008ne} for
the definition) has a same level as that for linear regression
($R^2=44.03\%$) for the range, the estimation of which gives the
estimation, $\alpha=1.037(\pm0.003)$ (shown by a straight line in
Fig.~\ref{fig:fig1}).

Similarly, for the range of $10^{4.5}<Y<10^{7.0}$, we have another
relation, namely
\begin{equation}
E(\ell|y)=\beta\,y+\text{const.},
\label{elybeta}
\end{equation}
with a constant $\beta$. The validity for this relation is checked
by the nonparametric ($R^2=47.18\%$) and linear ($R^2=47.09\%$)
regressions, the latter of which gives the estimation,
$\beta=0.655(\pm0.002)$.

\figpw{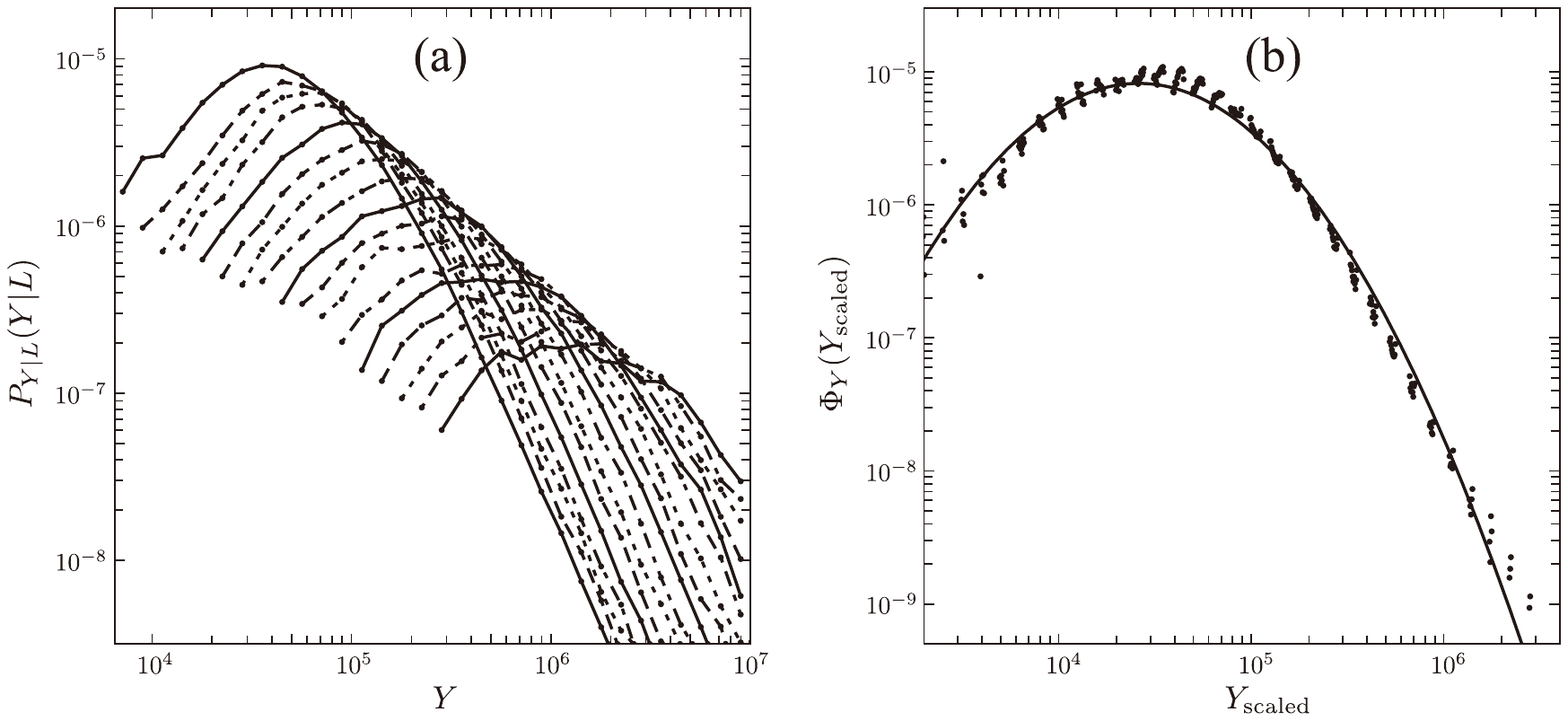}{%
  (a) The conditional PDF $\pyl(Y|L)$ for $L\in[5,200]$, where the
  conditioning values of $L$ are chosen at a logarithmically equal
  interval. 
  (b) The scaled conditional PDF $\Phi_Y(Y_{\rm scaled})$ defined by 
  Eq.~(\ref{s1}).
  Dots are the scaled data points and the curve is the lognormal function 
  given by Eq.~(\ref{f1}).}
  {0.85}{199 206 688 428}

We find that these relations are simple consequences from two
{\it scaling relations\/} for the conditional PDFs, $\pyl(Y|L)$ and
$\ply(L|Y)$. Fig.~\ref{fig:fig2}~(a) depicts the conditional PDF,
$\pyl(Y|L)$, with the conditioning values of $L$ are chosen at a
logarithmically equal interval corresponding to the range
$10^{0.7}<L<10^{2.0}$ in terms of histograms. 
By using the values of $\alpha$ estimated above, we find
that the conditional PDF obeys a scaling relation:
\begin{equation}
\pyl(Y|L)=\left(\frac{L}{L_0}\right)^{-\alpha}\!
\Phi_Y\!\left(Y_{\rm scaled}\right),
\label{s1}
\end{equation}
where $Y_{\rm scaled}:=(L/L_0)^{-\alpha}Y$ and
$\Phi_Y(\cdot)$ is a scaling function, as shown by the fact
that the PDFs $\pyl(Y|L)$ for different values of $L$ fall onto a
curve depicted in Fig.~\ref{fig:fig2}~(b). It is straightforward to
show that Eq.~(\ref{eylalpha}) follows from Eq.~(\ref{s1}).

\figp{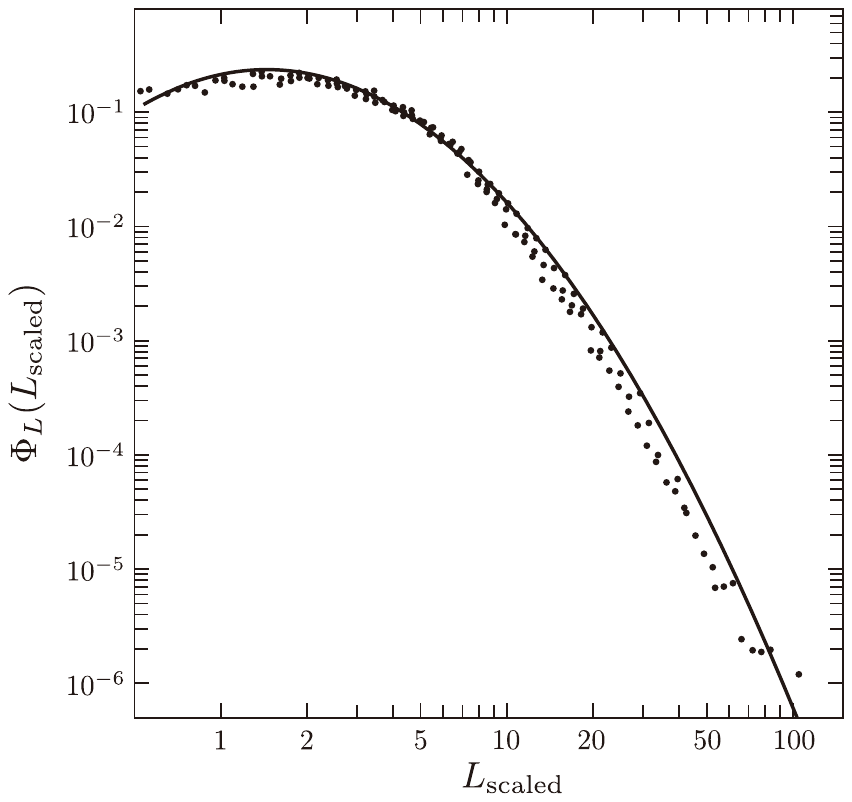}{%
  The scaled conditional PDF $\Phi_L(L_{\rm scaled})$ defined by 
  Eq.~(\ref{s2}).
  The curve is the lognormal function given by Eq.~(\ref{f4}).}
  {0.4}{456.042 202.375 696.737 428}

Similarly, we have another scaling relation for 
\begin{equation}
\ply(L|Y)=\left(\frac{Y}{Y_0}\right)^{-\beta}\!
\Phi_L\!\left(L_{\rm scaled}\right),
\label{s2}
\end{equation}
where $L_{\rm scaled}:=(Y/Y_0)^{-\beta}L$ and
$\Phi_L(\cdot)$ is a scaling function, as shown by
Fig.~\ref{fig:fig3}. And also Eq.~(\ref{elybeta}) immediately
follows from Eq.~(\ref{s2}).

The two scaling laws, Eqs.~(\ref{s1}) and (\ref{s2}),
which we collectively call {\it the double scaling law} (DSL)
have strong consequences to the function form of the joint PDF.
Let us briefly describe them in the following.

Let us choose the reference scales $Y_0$ and $L_0$
to be within the region of the $(Y,L)$-plane where DSL is valid.
Then by substituting $Y=Y_0$, $L=L_0$ into Eqs.~(\ref{pjoint1}), (\ref{s1}) and (\ref{s2}),
We obtain the marginal PDFs, $P_Y$ and $P_L$ in terms of
the invariant functions, $\Phi_Y$ and $\Phi_L$:
\begin{align}
P_L(L)&=\left(\frac{L}{L_0}\right)^\alpha 
\frac{\Phi_L( L)}{\Phi_L( L_0)}
\frac{\Phi_Y(Y_0)}{\Phi_Y\left((L/L_0)^{-\alpha} Y_0\right)}P_L(L_0), 
\label{pp1}\\
P_Y(Y)&=\left(\frac{Y}{Y_0}\right)^\beta
\frac{\Phi_Y(Y)}{\Phi_Y(Y_0)}
\frac{\Phi_L(L_0)}{\Phi_L\left((Y/Y_0)^{-\beta} L_0\right)}P_Y(Y_0).
\label{pp2}
\end{align}
From Eqs.~(\ref{s1}) and (\ref{s2}) and the above,
we arrive at the following equation for the $\Phi$s:
\begin{equation}
\frac{\Phi_L\left((Y/Y_0)^{-\beta}L\right)}{\Phi_L\left((Y/Y_0)^{-\beta}L_0\right)}
\frac{\Phi_L(L_0)}{\Phi_L(L)}
=\frac{\Phi_Y\left((L/L_0)^{-\alpha}Y\right)}{\Phi_Y\left((L/L_0)^{-\alpha}Y_0\right)}
\frac{\Phi_Y(Y_0)}{\Phi_Y(Y)}.
\label{solveme}
\end{equation}
This equation puts strong constraints the form of $\Phi$'s. 
We have converted the above to differential equations and have derived
{\it complete solutions} of Eq.~(\ref{solveme}).\footnote{Since the proof is too lengthy for this letter, it will be published elsewhere in near future.}
Depending on whether $\alpha\beta=1$ or not, the solution is qualitatively different, which we shall explain below.

When $\alpha\beta=1$, we find the following relation between $\Phi$'s is
necessary and sufficient for Eq.~(\ref{solveme}):
\begin{equation}
\Phi_L(L)=\Phi_Y\left((L/L_0)^{-\alpha}Y_0\right)
\left(\frac{L}{L_0}\right)^a 
\frac{\Phi_L(L_0)}{\Phi_Y(Y_0)}.
\end{equation}
In other words, we have one arbitrary function in the solution.
In this case, Eqs.~(\ref{pp1}) and (\ref{pp2}) implies that
\begin{align}
P_L(L) &=\left(\frac{L}{L_0}\right)^{-\mu_L-1}\!P_L(L_0),\label{ab1b}\\
P_Y(Y)&=\left(\frac{Y}{Y_0}\right)^{-\mu_Y-1}\!P_Y(Y_0),\label{ab1a}
\end{align}
with
\begin{equation}
\alpha=\frac{\mu_L}{\mu_Y}, \quad 
\beta=\frac{\mu_Y}{\mu_L},\quad
a=-\frac{\mu_L+\mu_Y+\mu_L\mu_Y}{\mu_Y}.\label{aba}
\end{equation}
This result is straightforward to understand: Due to $\alpha\beta=1$, we have
\begin{equation}
L_{\rm scaled}^{-\alpha}\propto Y L^{-\alpha} \propto Y_{\rm scaled}.
\end{equation}
Therefore, an arbitrary function of $Y_{\rm scaled}$ is 
a function of $L_{\rm scaled}$ as far as dependence on $Y$ and $L$ is concerned. This is why an arbitrary function is left in the solution. Also the marginal 
PDFs in Eqs.~(\ref{ab1b}) and (\ref{ab1a}) results from the relation (\ref{pjoint1}).

For $\alpha\beta\ne1$, we obtain,
\begin{align}
\Phi_Y(Y)&=e^{-\beta p y^2  + q y}\Phi_Y(Y_0),\label{f1}\\
\Phi_L(L)&=e^{-\alpha p \ell^2 + s \ell}\Phi_L(L_0),\label{f2}\\
P_L(L)&=e^{-\alpha(1-\alpha\beta) p \ell^2 + (s+(q+1)\alpha)\ell}P_L(L_0).\label{f4}\\
P_Y(Y)&=e^{-\beta(1-\alpha\beta) p y^2 + (q+(s+1)\beta)y}P_Y(Y_0).\label{f3}
\end{align}
The joint PDF is given by the following:
\begin{equation}
\PYL(Y,L)=e^{-\alpha p \ell^2+2\alpha\beta p\ell y-\beta p y^2
+s\ell + qy}\PYL(Y_0,L_0).
\label{jres}
\end{equation}
We find that in the limit $\alpha\beta\rightarrow1$ we obtain the power laws
for $P_L(L)$ and $P_Y(Y)$, which is consistent with the results 
(\ref{ab1b}) and (\ref{ab1a}) above.

The parameter of $p$ is estimated in two ways:
The best fit of the theoretical expression (\ref{f1}) in Fig.~\ref{fig:fig2}~(b) yields $p=0.692(\pm0.027)$,
while Eq.~(\ref{f1}) in Fig.~\ref{fig:fig3} yields $p=0.704(\pm0.016)$.
These measured values of $p$ agree with each other very well, with combined
value $p=0.698(\pm0.025)$, assuming equal weight and no correlation.
The marginal PDF $P_L(L)$ and $P_Y(Y)$ in Eqs.~(\ref{f4}) and (\ref{f3}) agrees with empirical data very well with these values of $p$.
These analysis show that our theoretical results above are in good agreement with data.

We stress that the above results are proven in the {\it local region} of the $(Y,L)$-plane where DSL is valid.
On the other hand, if the lognormal PDF (\ref{jres}) is valid 
{\it everywhere} on the $(Y,L)$ plane, one may obtain
the marginal PDFs and $\Phi$s as given in Eq.~(\ref{f2})--(\ref{f3}),
as $P_L(L)$ can be obtained by integrating Eq.~(\ref{jres}) over $Y \in [0,\infty)$, and then obtain  $\Phi_L(L)$ in Eq.~(\ref{f2}) from Eqs.~(\ref{s1}) and (\ref{s2}), and so forth, which constitute easy checks of the relation between various functions.

Let us now study the labor productivity $C:={Y}/{L}$ 
in case of $\alpha\beta\ne1$,  as such is the reality as shown empirically.
The joint PDF of $(C,L)$, $P_{CL}(C,L)$ is given by the following:
\begin{equation}
P_{CL}(C,L)=L\PYL(CL,L).\label{extraL}
\end{equation}
Substituting the expression (\ref{jres}) into the above,
we find that is express, we obtain that $P_{CL}(C,L)$ too is of lognormal
form like the r.h.s.\ of (\ref{jres}) with $\alpha, \beta, p$ replaced by
\newcommand\atilde{\tilde{\alpha}}
\newcommand\bt{\tilde{\beta}}
\newcommand\pt{\tilde{p}}
\begin{align}
\atilde:=\alpha-1, & \quad \bt:=\beta\frac{\alpha-1}{\alpha+\beta-2\alpha\beta},\\
\pt:=&\frac{\alpha+\beta-2\alpha\beta}{\alpha-1}\,p,
\label{deftilde}
\end{align}
respectively. By substituting empirical values found for
$\alpha, \beta$ and $p$ (the average of the two central values of $p$ found above),
we find that they are 
$\atilde=0.037(\pm 0.003)$, $\bt=0.072(\pm0.007)$, $\pt=6.353(\pm0.680)$.
By comparing this with Eq.~(\ref{f3}), we find the following marginal PDF for 
the productivity $C$:
\begin{align}
P_C(C)&\propto e^{-\gamma_1 c^2+\gamma_2 c},\label{pcc1}\\
\gamma_1&:=\frac{\alpha\beta(1-\alpha\beta)}{\alpha+\beta-2\alpha\beta}\,p,
\label{gscale}
\end{align}
where $c:=\ln(C/C_0)$ with $C_0:=Y_0/L_0$ and
the measured values of $\alpha, \beta, p$ yield $\gamma_1=0.456(\pm0.017)$.
Also, the conditional PDF $P_{L|C}(C,L)$ satisfies the scaling law;
\begin{align}
P_{L|C}(C,L)&=\left(\frac{C}{C_0}\right)^{-\bt}
\Psi_L\left(\left(\frac{C}{C_0}\right)^{-\bt}\!L\right),\\
\Psi_L&\propto e^{-\bt\pt c^2+\tilde{q}c},
\label{s3}
\end{align}
while the average of the labor $L$ for a given productivity $C$ is given by the following;
\begin{equation}
E[L|C]\propto \left(\frac{C}{C_0}\right)^{\bt}.
\label{elcbt}
\end{equation}

\figp{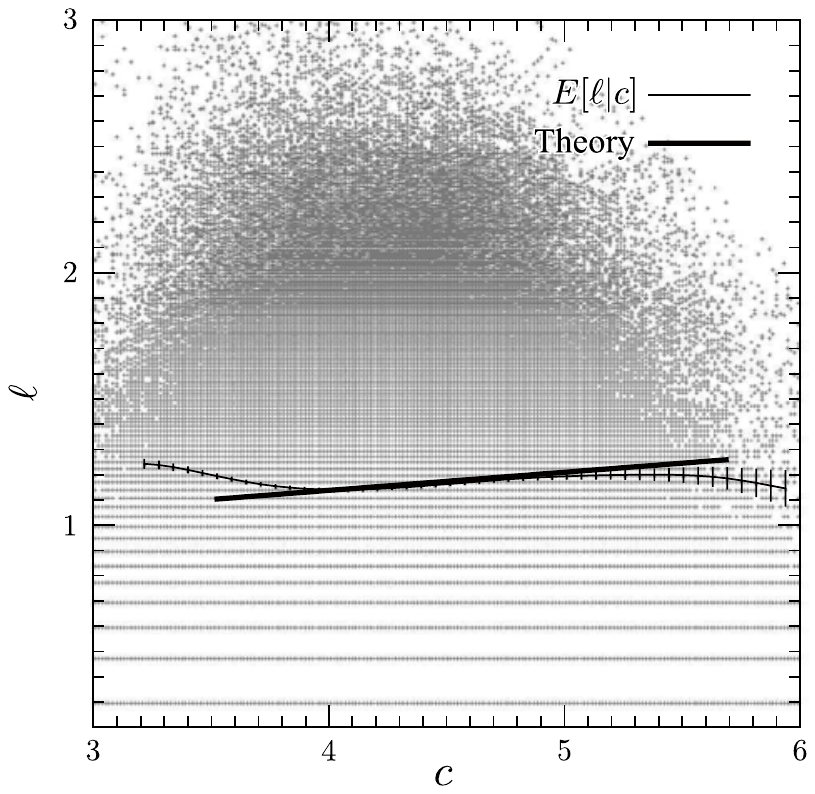}{%
  Scatter plot for $(C,L)$ (gray dots). The curve with error bars is
  the nonparametric estimation for $E(\ell|c):=E(\ln L|C)$, while the
  thick straight line is the theoretical prediction (\ref{elcbt})
  with $\bt$ given by Eq.~(\ref{deftilde}) (in units of $\ln 10$).}{0.4}
  {188 316 419 540}

\figp{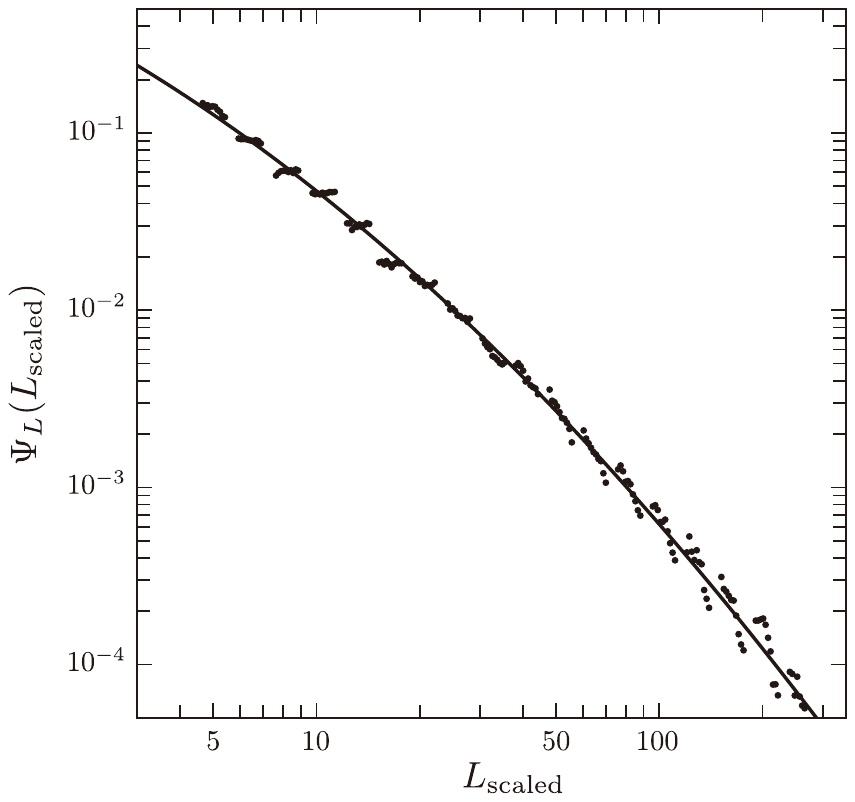}{%
  The scaled conditional PDF $\Psi_L(L_{\rm scaled})$ defined by 
  Eq.~(\ref{s3}). The curve is the lognormal function given in the
  same equation.}{0.4}{456 202 697 428}

Fig.~\ref{fig:fig4} checks the result of Eq.~(\ref{elcbt}) with
the parameter $\bt$ estimated from the relation in Eq.~(\ref{s3}),
where the thick straight line is the theoretical calculation. 
Fig.~\ref{fig:fig5} depicts the scaling function $\Psi(\cdot)$, where
the curve is given by the fitting in Eq.~(\ref{s3}). Both of these
results confirm our results in the region where the scaling relations
are valid.

In this paper, we have shown that the Japanese data for some one million firms
show that the firm distribution in $(Y,L)$ plane satisfy the double scaling law (DSL). We have shown that DSL leads to either power-law for the marginal PDF of $Y$ and $L$, or the lognormal PDF for the joint PDF of $Y$ and $L$.
Although we have concentrated on these specific two variables because of their importance for the labour productivity, we believe that other financial quantities obey DSL. Furthermore, joint PDF of more than two variables are expected to obey the similar scaling laws, which yield extensions of the results explained above, including multi-variate lognormal distributions, yielding simple relations such as Eq.~(\ref{gscale}) between scaling exponents. These laws offer straightforward guidance to the method of aggregation: the macroscopic quantities are now expressed in terms of just a few parameters in the PDFs, conditioned by DSL.
Thus scaling laws and the resulting lognormal distributions should be the basic ingredient of economics of HIA, which is an equivalence of the statistical physics, bridging micro and macro economy.

\begin{acknowledgments}
We would like to thank Hiroshi Iyetomi, Yuichi Ikeda, Wataru Souma and Hiroshi Yoshikawa for discussions, support and encouragements. We also would like to thank Mr.\ Shigeru Hikuma, the President of the CRD Association for providing us with their database.
The present work was supported in part by 
{\it the Program for Promoting Methodological Innovation in Humanities and Social Sciences, and by Cross-Disciplinary Fusing} of the Japan Society for the Promotion of Science, 
{\it Grant-in-Aid for Scientific Research (B)} 20330060 (2008-10),
and {\it Invitation Fellowship Program for Research in Japan (Short term)} ID No.S-09132 of the Ministry of Education, Science, Sports and Culture, Japan.
\end{acknowledgments}


%

\end{document}